\documentclass{appolb}
\usepackage{epsfig}
\begin{document}
\pagestyle{plain}
\eqsec  
\newcount\eLiNe\eLiNe=\inputlineno\advance\eLiNe by -1
\title{
\hfill{\small {\bf MKPH-T-02-03}}\\
Interaction of $\eta$ mesons with a three-nucleon system
\thanks{Supported by Deutsche Forschungsgemeinschaft SFB 443.}
\thanks{Contribution to MESON 2002, 7th Int.\ Workshop on
Production, Properties and Interaction of Mesons, Cracow, Poland, May
24-28, 2002.
}}
\author{
A.\ Fix and H.\ Arenh\"ovel
\address{Institut f\"ur Kernphysik, Johannes Gutenberg-Universit\"at,
D-55099 Mainz, Germany}}
\maketitle
\vspace*{-1cm}
\begin{abstract}
A study of the $\eta$-$3N$ interaction in the energy region 
near the $\eta$-$^3$H elastic 
scattering threshold is presented. 
The calculational scheme is based on 
the four-body scattering formalism. 
A manageable form of the integral equations 
is achieved by using a separable ansatz for the driving two-body 
$\eta N$- and $NN$-forces as well as for the subamplitudes appearing in 
the (1+3) and (2+2) partitions of the $\eta$-$3N$ system. 
Results presented for the $\eta$-$^3$H 
scattering length point to the existence of a
virtual (antibound) $s$-wave $\eta$-$^3$H state. As a consequence,
a large enhancement of the cross section for 
$\eta$-production from three-body nuclei is predicted.   
\end{abstract}
\vspace*{-.3cm}
\vspace*{.1cm}
The experimental results for $\eta$-production from the lightest nuclei
demonstrate a rather strong energy dependence in the region of a few MeV above
threshold \cite{May96,Cal97,Hej02}.
One observes a very rapid rise of the $\eta$-yields with
energy, so that the predictions of simple models,
neglecting any final state interaction, underestimate the data sizeably.
Some of the models interpret this striking
enhancement of the near-threshold $\eta$-production as an indirect
indication that the $\eta$-meson can form bound states already
with $s$-shell nuclei. Whether the $\eta$-nuclear 
interaction allows the existence of such objects remains one of the
most exciting but still unproven hypothesis of meson-nuclear physics. 
As for the $\eta$-$3N$ system, several methods have been
developed \cite{Wilk93,Trs97,Wyc95,Rakit96,Khem02}, where different
approximations were used to circumvent the difficulties associated
with a direct solution of the four-body problem. The crucial point of all
these approaches is the  
neglect of the target excitation during the interaction with the
$\eta$-meson, which may be too restrictive if few-nucleon targets are
considered \cite{FiAr01}. Therefore, our aim was to study  
the $\eta$-$3N$ interaction within the formally exact four-body scattering
theory. 

The formal basis of our calculation is described in detail in
\cite{FiAr02}. As a principal tool, we have employed the four-body 
integral equations which were reduced to the two-body form within the
quasiparticle approach \cite{GS67,AGS70}. 
As a starting two-body input, we have used a simple rank one separable
parametrization for the $\eta N$- and $NN$-amplitudes with
Hulth\a`en form factors. In the $\eta N$-channel only the excitation of
the $S_{11}(1535)$-resonance was taken into account. Also the $NN$
sector was restricted to the dominant $^1S_0$ and $^3S_1$ states. 
The separable representation for the integral kernels corresponding to
the (1+3)- and (2+2)-partitions of the $\eta$-$3N$ system was
obtained by using the Hilbert-Schmidt method. 
\vspace*{-.2cm}
\begin{table}[hbt]
\renewcommand{\arraystretch}{1.2}
\caption{\small Predictions for the $\eta\,^3$H scattering lengths
for given $\eta N$-inputs together with results 
of \protect{\cite{Wyc95,Rakit96}}.}
\begin{center}
\begin{tabular}{|crc|}
\hline
$a_{\eta N}$ [fm] & $a_{\eta\, ^3\mathrm{H}}$ [fm]
& $a_{\eta\, ^3\mathrm{H}}$ [fm] (this work)   \\
\hline
$0.57+i\,0.39$   &  $1.32+i\,4.37$ \protect{\cite{Wyc95}}
& $2.23+i\,3.00$ \\
$0.29+i\,0.36$   &  $0.58+i\,2.17$ \protect{\cite{Wyc95}}
& $0.97+i\,1.72$ \\
$0.27+i\,0.22$   &  $0.41+i\,2.00$ \protect{\cite{Rakit96}}
& $0.69+i\,0.67$ \\
$0.55+i\,0.30$   & $-1.56+i\,3.00$ \protect{\cite{Rakit96}}
& $2.35+i\,1.68$ \\
$0.75+i\,0.27$   &  & $4.19+i\,5.69$ \\
\hline
\end{tabular}
\label{tab1}
\end{center}
\end{table}
\vspace*{-.5cm}

In Table \ref{tab1}, our predictions for the $\eta\,^3$H
scattering length related to different $\eta N$ inputs 
are compared with those of \cite{Wyc95,Rakit96}. The
last result corresponds to an $\eta N$ interaction with the scattering length
of~\cite{Wyc97}, 
which is considered in modern analyses as the most realistic one. 
We have used this value in the calculations presented below. 
The main conclusion, we can drawn, is that within our model 
the $\eta$-$3N$ interaction turns out to be not strong enough to bind 
this system. This inference does not support the results of the
finite-rank-approximation \cite{Rakit96} as well as those of the simple optical
model \cite{Wilk93,Trs97}, where a bound state appears already   
for rather modest values of the $\eta N$ scattering length. 
On the other hand, the virtual (antibound) state, generated
by the strong attraction in the $\eta$-$3N$ system, 
must substantially influence
its dynamical properties. As one can see, the last prediction for $a_{\eta 
^3\mathrm{H}}$ in Table~\ref{tab1} is impressively large, indicating that a 
virtual pole lies very near
zero energy. Therefore, a strong effect of the final state interaction
is to be expected in the $\eta$-production reactions. 

To confirm this conclusion, we present below our results for the
production of $\eta$-mesons by pions and photons from three-body
nuclei. These
reactions were already investigated within the frame of more
approximate models (see e.g.\ \cite{Liu92,Kamal93,Shev02}). 
As noted above, our calculations were performed with the $\eta N$-parameters,
corresponding to $a_{\eta N}$ = (0.75 + i\,0.27) fm (see Table
\ref{tab1}). For the 
isoscalar part $T_s$ of the photoexcitation amplitude $T_{\gamma N\to S_{11}}$
we have employed (see e.g. \cite{Hej99}) 
$T_s/T_{\gamma p\to S_{11}}=0.1$.
\begin{figure}[htb]
\centerline{\epsfxsize=11cm \epsffile{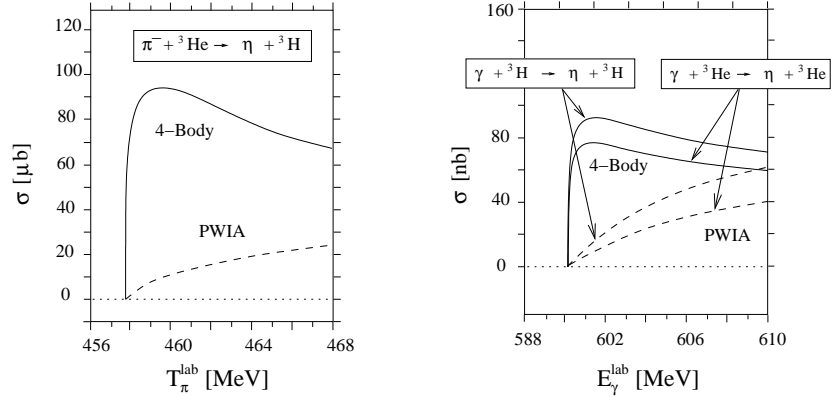}}
\vspace*{-.2cm}
\caption{\small Total cross section for pion (left) and photon (right)
induced $\eta$-production.} 
\label{fig3}
\end{figure}
\begin{figure}[htb]
\centerline{\epsfxsize=11cm \epsffile{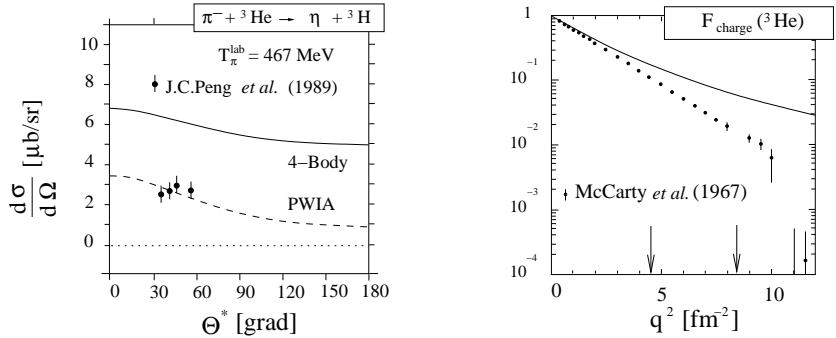}}
\vspace*{-.2cm}
\caption{\small Left panel: differential cross section for
$^3$He$(\pi^-,\eta)^3$H at pion kinetic energy 467 MeV 
without (PWIA) and with (4-Body) final state interaction. 
Data from \protect{\cite{Peng89}}.
Right panel: charge form factor of $^3$He predicted by our
calculation with a Yamaguchi potential. Arrows
indicate the region of the transferred momenta, associated with the
angular distribution shown on the left panel. 
Data from \protect{\cite{McCart77}}.
} 
\label{fig4}
\end{figure}
Final state interaction has been included in the most important $s$-wave part.
Our predictions for the total and differential cross sections are presented in
Figs.\ \ref{fig3}, \ref{fig4}.
As is expected, the $\eta$-$3N$ virtual state results in a
very strong enhancement of the $\eta$-yields near the production threshold.
Furthermore, due to the increase of the $s$-wave part of the scattering
amplitude, the differential cross section exhibits a rather weak angular
dependence. At the same time, we are not able to reproduce the
experimental 
$^3$He$(\pi^{-},\eta)^3$H cross section \cite{Peng89}. As may be
seen in Fig.\,\ref{fig4} our results strongly overestimate 
the data. In essence, this is due to the
oversimplified treatment of the nuclear sector. In order to illustrate this, 
we show in Fig.~\ref{fig4}
our predictions for the $^3$He electromagnetic charge form factor.  
One can see, that in the relevant region of momentum transfers, 
characteristic for 
$\eta$-production, the calculation noticeably overestimates the corresponding 
experimental values \cite{McCart77}. In fact, 
it is well known that the simple 
Yamaguchi $NN$-potential, used in the present calculation, has 
too much attraction at high momenta, with the effect 
that the nucleons
in the target are too close to each other. This problem,
of course, may be solved by using more refined $NN$-potentials, where also
the short range repulsion is taken into account. 

In summary, the investigation of the $\eta$-$3N$ interaction in the
frame of the 4-body scattering theory shows that within the range of the 
$\eta N$-interaction strength, approaching realistic values, 
this system possesses a 
virtual pole located rather near the elastic scattering threshold. Its
existence manifests itself in a striking enhancement of the
$\eta$-production from three-body nuclei. This conclusion
agrees qualitatively with the available experimental results
\cite{May96}. On the other hand, our calculation fails in describing the
data for the reaction $^3$He$(\pi^{-},\eta)^3$H \cite{Peng89},
which must be attributed to the oversimplified 
treatment of the nuclear wave function in the region of large
momentum transfers. Thus for any realistic description of the data, an
essential refinement of the nuclear sector is needed.

\end{document}